\documentstyle[12pt]{article}
\begin{document}
\begin{center}
{\bf Coding Theorems for Quantum Communication Channels}
\vskip10pt
A. S. Holevo\\
Steklov Mathematical Institute, Moscow
\end{center}

\vskip30pt
\centerline{\sc I. The capacity and the entropy bound}
\vskip10pt\

Let $\cal H$ be a
Hilbert space providing a quantum-mechanical description for
the physical carrier of information. A simple model of quantum
communication channel  consists of the input alphabet $A = \{1,...,a\}$ and a mapping
$i\rightarrow S_i$ from the input alphabet to the set of quantum states in
$\cal H$. A quantum state is a {\sl density operator}, i. e.  positive
operator $S$ in $\cal H$ with unit trace, Tr$S = 1$.  Sending a letter
$i$ results in producing the signal state $S_i$ of the information carrier.

Like in the classical case, the input is described by an {\sl apriori
probability distribution} $\pi = \{\pi_i\}$ on $A$. At the receiving
end of the channel a
quantum measurement  is
 performed, which mathematically is described by a
{\sl resolution of identity} in $\cal H$, that is by
 a family $X = \{ X_j\}$ of positive
operators in $\cal H$ satisfying $\sum_j X_j = I$, where $I$ is the unit
operator in $\cal H$ \cite{xhol73}.
The probability of the output $j$ conditioned upon the
input $i$ by definition is equal to $P(j|i) = \mbox{Tr}S_i X_j$.
The classical case is embedded into this picture by assuming that all operators
in question commute, hence are diagonal in some basis labelled by index
$\omega$; in fact by taking $S_i = \mbox{diag}[ S(\omega |i) ], X_j
= \mbox{diag}[ X(j |\omega) ]$, we have a classical channel with transition
probabilities $S(\omega |i)$ and the classical
decision rule $X(j |\omega)$, so that $P(j|i) = \sum_{\omega}X(j |\omega)
S(\omega |i)$. We call such channel {\sl quasiclassical}.

The Shannon information is
given by the usual formula $$I_1 (\pi , X ) = \sum_j \sum_i \pi_i
P(j|i)\mbox{log}\left( \frac{P(j|i)} {\sum_k \pi_k P(j|k)}\right) .\eqno(1)$$
Denoting by $H(S) = - \mbox{Tr}S\mbox{log}S$ the von Neumann entropy of a
state  $S$,  we {\sl assume} that $H(S_i) < \infty$.
If $\pi = \{\pi_i \}$ is an apriori distribution on $A$, we denote
$${\bar S}_{\pi} = \sum_{i\in A}\pi_i S_i ,\qquad {\bar H}_{\pi}
(S_{(\cdot )}) = \sum_{i\in A}\pi_i H(S_i )$$
and $$\Delta H (\pi ) = H({\bar S}_{\pi}) -
 {\bar H}_{\pi}(S_{(\cdot )}).$$ 

The famous {\sl quantum entropy bound} says that
$$\sup_{X}I_1 (\pi, X)\leq \Delta H (\pi ), \eqno(2)$$
with the equality achieved if and only if all the operators $\pi_i S_i$
commute. The inequality was explicitly conjectured in \cite{xgordon},
and discussed in \cite{xforney},
\cite{xlev} and elsewhere in the context of
conventional quantum measurement theory.
The first published proof appeared in \cite{xhol73}. It is
worthwhile to mention also that
 this bound is closely related to the fundamental property of decrease of
quantum relative entropy under completely positive maps
developed in \cite{xlin} and in
\cite{xuhl} (see \cite{xyuen} for history and some
generalizations of the entropy bound).

 In the same way we can consider the product channel in the tensor product
Hilbert space ${\cal H}^{\otimes n} =
{\cal H}\otimes ...\otimes {\cal H}$ with the input alphabet $A^n$ consisting
of words $w = (i_1 ,...,i_n )$ of length $n$, and with the density operator
$S_w = S_{i_1}\otimes ...\otimes S_{i_n}$ corresponding to the word $w$.
If ${\pi}$ is an apriori distribution on $A^n$ and $X$ is a resolution of
identity in ${\cal H}^{\otimes n}$, we define the information quantity $I_n
(\pi , X )$ by the formula similar to (1).  Denoting $C_n = \sup_{\pi , X} I_n
( \pi ,X),$ we have the property of superadditivity $C_n + C_m \leq C_{n+m}$,
hence the following limit exists $$C = \lim_{n \to \infty}C_n /n ,$$
which is called the {\sl capacity} of the initial channel \cite{xhol79}. This
definition is justified by the fact easily deduced from the classical Shannon's
coding theorem, that $C$ is the least upper bound of the rate (bits/symbol) of
information which can be transmitted with asymptotically vanishing error.  More
precisely, we call by {\sl code $(W, X)$
of size} $M$ a sequence $(w^1 , X_1 ),..., (w^M
, X_M )$, where $w^k$ are words of length $n$, and $\{ X_k \}$ is a family of
positive operators in ${\cal H}^{\otimes n}$, satisfying $\sum_{k=1}^M X_k \leq
I$. Defining $X_0 = I - \sum_{k=1}^M X_k$, we have a resolution of identity in
${\cal H}^{\otimes n}$.
The average error probability for such a code is $${\bar \lambda}
(W, X) =
\frac{1}{M} \sum_{k=1}^M [1 - \mbox{Tr}S_{w^k} X_k ].\eqno(3) $$
Let us denote $p(M,n)$ the minimum of this error probability with respect to all codes of the size
$M$ with words of length $n$.  Then  $p(2^{n(C - \delta )},n) \rightarrow
0\qquad \mbox{and} \qquad p(n, 2^{n(C + \delta )}) \not\rightarrow 0,
$ where $\delta > 0,$ as $  n \rightarrow \infty$.

Applied to $I_n (\pi, X)$ and combined with the
additivity  and continuity properties of $\Delta H (\pi )$  the entropy bound (2)
implies $C \leq \max_{\pi}\Delta H(\pi) \equiv {\bar C}.$ Thus
$$C_1 \leq C \leq {\bar C}.$$ For a classical channel $C_n = n C_1$, and all the
three quantities coincide. A striking feature of quantum case is possibility
of the inequality $C_1 < C$ implying strict superadditivity of the
information quantities $C_n$ \cite{xhol79}. In a sense, there is a kind of
``quantum memory'' in channels, which are the analog of classical memoryless
channels. This fact is  just another manifestation of the ``quantum
nonseparability'', and in a sense is
dual to the existence of Einstein - Podolsky -
Rosen correlations: the latter
are due to entangled (non-factorizable)
states and hold for disentangled measurements while the
superadditivity is due to entangled measurements and holds for
disentangled states.

The inequality $C\not= C_1$ raised the problem of the
actual value of the capacity $C$. A possible conjecture was $C = {\bar C}$,
but the proof for it came only recently, first for the pure state (noiseless)
channels in the paper of Hausladen, Jozsa, Schumacher, Westmoreland and Wootters
\cite{xjozsa}, and then for the case of arbitrary signal states in \cite{xhol} 
and in \\ 
\cite{xwest}. Since the entropy bound (2) and the classical weak converse
provide the converse of the quantum coding theorem,
the main problem was the proof of the direct coding theorem, i. e. of the
inequality $C\geq {\bar C}$.
\vskip20pt
\centerline{\sc II. The pure state channel}
\vskip10pt
Following Dirac's formalism, we shall denote vectors of $\cal H$ as
$ |\psi>$,  and hermitean conjugate vector of the dual space -- as $<\psi |$.
 Then $<\phi |\psi >$ is the inner product of $|\phi>, |\psi>$
and $|\psi><\phi |$ is the outer product, i. e. operator $A$ of rank 1,
acting on vector $|\chi>$ as $A |\chi> = |\psi><\phi |\chi>.$
If $|\psi>$  is a unit vector, then $|\psi><\psi |$
is the orthogonal projection onto
$|\psi>$. This is a special density operator, representing {\sl pure}
state of the system. Pure states are precisely extreme points of the convex
set of all states; an arbitrary state can be represented as a mixture
of pure states, i. e. by imposing classical randomness on pure states.
In this sense pure states are ``noiseless'', i. e. they contain no classical
source of randomness.

Let us consider a {\sl pure state channel} with
 $S_i = |\psi_i\!\!><\psi_i|$ . Since the entropy of a pure state is
 zero, $\Delta H (\pi ) = H( {\bar S}_{\pi})$ for such a channel.
If a decision rule $X =\{X_j \}$ is applied at the output then
$P(j|i) = <\psi_i |X_j \psi_i>.$
A system $\{|\phi_j>\}$ of (unnormalized) vectors
is called {\sl overcomplete} if
$\sum_j |\phi_j><\phi_j | = I$.
Every overcomplete system
gives rise to the decision rule $X_j = |\phi_j><\phi_j |$ for which
$P(j|i) = |<\psi_i |\phi_j >|^2.$

The first step in getting a lower bound for the capacity
$C$ has geometric nature
and amounts to obtaining a tractable upper bound for the
average error (3) minimized over all decision rules.
Sending a word $w = (i_{1},\ldots,i_{n})$
produces the tensor product vector $\,\psi_{w} =
\psi_{i_{1}} \otimes \ldots \otimes \psi_{i_{n}} \in {\cal H}^{\otimes n}$.
Let $(W, X)$ be a code of size $M$. Let us restrict for a while to the
subspace of ${\cal H}^{\otimes n}$ generated by the vectors
$\psi_{w^1},\ldots,\psi_{w^M}$, and consider the Gram matrix
$\,\Gamma (W) = [<\psi_{w^i}|\psi_{w^j}>]\,$ and the
Gram operator $G(W) =
\sum_{k=1}^{M}|\psi_{w^k}><\psi_{w^k}|$.
This operator has the matrix $\Gamma (W)$ with respect to the
overcomplete system
$$
|\hat{\psi}_{w^k}> = G(W)^{-1/2} |{\psi}_{w^k}>\,;\,
\quad k=1,\ldots,M\;.\eqno(4)
$$
The resolution of identity of
the form $$X_{k} = |\hat{\psi}_{k}><\hat{\psi}_{k}|\eqno(5)$$ approximates
the quantum maximum likelihood decision rule (which in general
cannot be found explicitly);
the necessary normalizing factor $G(W)^{-1/2}$ is a major source of
analytical difficulties in the noncommutative case. Note that the vectors
$\psi_{w^1},\ldots,\psi_{w^M}$ need not be linearly independent; in the case
of linearly independent coherent state vectors (5) is related to the
``suboptimal receiver'' described in \cite{xhel}, Sec. VI.3(e).
It was shown in
\cite{xhol78} that by using this decision rule one obtains the upper bound
$$
\min_{X} \bar{\lambda}(X, W) \leq {2 \over M}\,{\rm Sp}
\left(E - \Gamma (W)^{1/2}\right) = {1 \over M}\,{\rm Sp}
\left(E - \Gamma (W)^{1/2}\right)^2 ,
\eqno(6)$$
where $E$ is the unit $M\times M$-matrix and $\rm Sp$ is the trace of
$M\times M$-matrix.
This bound is ``tight'' in the sense that there is a similar lower bound.
However it is difficult to use because of the presence of
square root of the Gram matrix. A
simpler but coarser bound is obtained by using the operator inequality
$(E - \Gamma (W)^{1/2})^2 \leq (E - \Gamma (W))^2$:
$$
\min_{X} \bar{\lambda}(W, X) \leq {1 \over M}\,{\rm Sp}
\left(E - \Gamma (W)\right)^2 = {1 \over M}\,{\rm Tr}
\sum\sum_{r\not= s} S_{w^r} S_{w^s}.
\eqno(7)$$
As shown in \cite{xhol78}, this bound is asymptotically
equivalent (up to the factor 1/4)
to the tight bound (6) in the limit of ``almost orthogonal''
states $\Gamma (W)\rightarrow E$. On the other hand,
different words are ``decoupled'' in (7) which makes it suitable for application
of the random coding.

Just as in the classical case, we assume that the words $w^1 ,..., w^M$ are
chosen at random, independently and with the probability distribution
$${\sf P}\{w =( i_1,\ldots,i_n)\} = \pi_1 \ldots \pi_n . \eqno (8)$$
Then for each word $w$ the expectation
$${\sf E}\,  S_w = {\bar S}_{\pi}^{\otimes n},\eqno(9)$$
and by taking the expectation of the coarse bound (7) we obtain,
due to the independence of $w^r, w^s$
$$p (M, n)\leq {\sf E} \min_{X}\bar{\lambda}(W, X) \leq (M-1) {\rm Tr}
({\bar S}_{\pi}^{\otimes n})^2 = (M-1)2^{-n\log {\rm Tr}{\bar S}_{\pi}^2}.$$
By denoting $${\tilde C} = -\log\min_{\pi}{\rm Tr}{\bar S}_{\pi}^2
= -\log\min_{\pi}\sum_{i,j}\pi_i\pi_j |<\psi_i |\psi_j >|^2,\eqno(10)$$
we conclude that $C\geq {\tilde C}$ . There are cases (e. g.
pure state binary channel) where ${\tilde C} > C_1$, so this is sufficient
to establish $C > C_1$, and hence the strict superadditivity of $C_n$
\cite{xhol79}, but not sufficient
to prove the coding theorem, since ${\tilde C} <{\bar C}$ unless
the channel is quasiclassical. A detailed comparison of the quantities
$C_1 , {\bar C}$ for different quantum channels was made by
Ban, Hirota, Kato, Osaki and Suzuki \cite{xhirota}. The quantity $\tilde C$ was discussed in
\cite{xhol79}, \cite{xstr}, but its real information theoretic meaning
is elucidated only in connection with the quantum reliability
function  (see (15) below). 

The proof of the inequality $C\geq {\bar C}$ given in \cite{xjozsa}
achieves the goal by using
the approximate maximum likelihood improved with projection onto the ``typical subspace''
of the density operator ${\bar S}_{\pi}^{\otimes n}$ and the correspondingly
modified
coarse bound for the error probability. The coarseness of the bound is thus
compensated by eliminating ``non-typical'' (and hence far from being orthogonal)
components of the signal state vectors. More precisely,
let us fix small positive $\delta$, and let
$\lambda_j$ be the eigenvalues, $|e_j >$  the eigenvectors
of ${\bar S}_{\pi}$.
Then the eigenvalues and eigenvectors of ${\bar S}_{\pi}^{\otimes n}$ are
$\lambda_J = \lambda_{j_1} \cdot ... \cdot
\lambda_{j_n},\quad |e_J > = |e_{j_1}>\otimes ...\otimes |e_{j_n}>$
where $J = (j_1,...,j_n )$. The spectral projector onto the {\sl typical subspace}
is defined as $$P= \sum_{J\in B} |e_J ><e_J |,
$$ where $B = \{J: 2^{-n[H({\bar S}_{\pi})+\delta ]} < \lambda_J <
2^{-n[H({\bar S}_{\pi})-\delta ]}\}$. This concept plays a central role
in ``quantum data compression'' \cite{xschum}. In a more
mathematical context a similar notion
appeared in \cite{xpetz}, Theorem 1.18. Its application to the present
problem relies upon the following two basic properties: first, by definition,
$$\|{\bar S}_{\pi}^{\otimes n} P\| < 2^{-n[H({\bar S}_{\pi})-\delta ]}.
\eqno(11)$$
Second, for fixed small positive $\epsilon$ and large enough $n$
$$\mbox{Tr}{\bar S}_{\pi}^{\otimes n}(I - P)\leq\epsilon ,   \eqno(12)$$
because a sequence $J\in B$ is
typical for the probability distribution given by eigenvalues
$\lambda_J$ in the sense of classical
information theory  \cite{xgal}, \cite{xinf}.

By replacing the signal state vectors $|\psi_{w^k}>$ with unnormalized
vectors $|{\tilde \psi}_{w^k}>=P|\psi_{w^k}>$, defining the corresponding
approximate maximum likelihood decision rule,and denoting ${\tilde \Gamma (W)}$
the corresponding Gram matrix, the modified upper bound
$$
\min_{X} \bar{\lambda}(W, X) \leq {1 \over M}\,\{{\rm Sp}
\left(E - {\tilde \Gamma} (W)\right) +
{\rm Sp}
\left(E - {\tilde \Gamma} (W)\right)^2\}$$ $$= {1 \over M}\,
\sum_r \{ {\rm Tr} S_{w^r}(I-P) +
\sum_{s\not= r} {\rm Tr}S_{w^r} P S_{w^s} P\}$$
is obtained in \cite{xjozsa}. Applying the random coding and using (9) and
the properties (11), (12) of the typical subspace, one gets for large $n$
$$p(M, n)\leq\{\epsilon + (M-1)2^{-n[H({\bar S}_{\pi}) - \delta ]}\},$$
resulting in the inequality $C\geq {\bar C}$.

It is known, however, that
in classical information theory the coding theorem can be proved
without resorting to typical sequences, by mere use of clever estimates
for the error probability \cite{xgal}. Moreover, in this way one obtains
the exponential rate of convergence for the error probability, the
so called reliability function
$$E(R) = \lim_{n \to \infty}
\sup {1 \over n}\log {1 \over p(\mbox{e}^{nR},n)} \;,\quad 0 < R < C \;.$$
This puts us onto the idea of  trying to apply the random coding procedure
directly to the tight bound (6) in the quantum case. This is realized
in \cite{xbur}.
Rather remarkably, the expectation
can be calculated explicitly, 
$${2 \over M}\,{\sf E}\,{\rm Sp}\left(E - \Gamma (W)^{1/2}
\right) = {\rm Tr}\,f({\bar S}_{\pi}^{\otimes n}),$$
where $$
f(z)={2 \over M}[1-\sqrt{1+(M-1)z}+(M-1)(1-\sqrt{1-z})].$$
This function strangely resembles the expression
for the Bayes error in the ``equiangular'' case \cite{xhel} rel. (VI.2.10), 
although does
not coincide with it. The function $f(z)$ admits standard estimates
$$
f(z)\leq z \min\,\left\{(M-1)z, 2\right\}\leq 2 (M-1)^{s}
z^{1+s}\,,\; 0\leq s \leq 1\,,$$
allowing to prove the following result \cite{xbur}

{\bf Theorem 1}. {\sl For all $M,n$ and} $0\leq s\leq 1$
$${\sf E}\,\min_{X}\bar{\lambda}(W, X) \leq
2(M-1)^{s}\left[{\rm Tr}\,{\bar S}_{\pi}^{1+s} \right]^{n}. \eqno(14)$$

It is natural to introduce the function $\mu(\pi,s)$ similar to
analogous function in classical information theory
\cite{xgal}, Sec. 5.6
$$
\mu (\pi,s) =
- \log {\rm Tr}\,{\bar S}_{\pi}^{1+s} =
- \log \sum_j \lambda_j^{1+s}.$$ Then
$$E(R) \geq
\max_{\pi} \max_{0\leq s \leq 1} \left(\mu(\pi,s)-s R\right)\,
\equiv E_r (R).$$
On the other hand, it appears possible to apply in the quantum case the
``expurgation'' technique from \cite{xgal},
Sec. 5.7, resulting in the bound
$$ E(R)\geq \max_{\pi}\max_{s \geq 1} ({\tilde \mu}(\pi, s) -
s R )\equiv E_{ex}(R) ,$$ where
$${\tilde \mu}(\pi, s) = - s \ln
\sum_{i,k }\pi_i \pi_k |<\psi_i |\psi_k>|^{2 \over s}.$$
The behavior of the lower bounds $E_r (R), E_{ex} (R)$ can be studied by
the methods of classical information theory, see \cite{xbur}.
In particular, it follows easily that $C\geq \max_{\pi}\mu'(\pi , 0) = {\bar C}$.
Thus the rate $C-\delta$ can be attained with the approximate 
maximum likelihood decision
rule (5), (4) without projecting onto the typical subspace.

We also remark that
$${\tilde \mu}(\pi, 1) = \mu (\pi, 1) = - \log\mbox{Tr}{\bar
S}_{\pi}^2, \eqno(15)$$ and that the common linear portion of the functions
$E_r (R), E_{ex} (R)$ is just $\mu (\pi, 1) - R$.
\vskip20pt
\centerline{\sc III. General signal states with finite entropy}
\vskip10pt
The general case is substantially more complicated
already on the level of quantum Bayes problem; in particular, so far no
upper bound for the average error probability is known, generalizing
appropriately the
geometrically simple bound (6). The proof given in \cite{xhol}
is based rather on a noncommutative generalization
of the idea of ``jointly typical'' sequences in classical theory \cite{xinf}.
This is realized by substituting in the average error probability (3) the
decision rule
 $$X_{w^k} = (\sum_{l=1}^M PP_{w^l}P)^{-\frac{1}{2}}PP_{w^k} P
(\sum_{l=1}^M PP_{w^l}P)^{-\frac{1}{2}},  \eqno(16)$$ where $P_{w^k}$ is
a proper generalization of the typical projection for the density operators
$S_{w^k}$ . The essential properties of $P_{w^l}$ are
$$P_{w^k}\leq S_{w^k} 2^{n[{\bar H}_{\pi} (S_{(\cdot)}) + \delta ]}, \eqno(17)$$
$${\sf E}\mbox{Tr}S_{w^k} (I - P_{w^k} )\leq\epsilon,\eqno(18)$$

After substituting (16) into (3) and performing a number of rather
laborious steps intended to
get rid of the normalization factors in (16) and thus to obtain an expression
in which the different words are ``decoupled'', one arrives at the estimate
$$\min_{X}{\bar \lambda}(W, X)
\leq \frac{1}{M}\sum_{k=1}^M \{ 3\mbox{Tr}S_{w^k} (I - P) +
\sum_{ l\not= k} \mbox{Tr}P S_{w^k} P P_{w^l}
+ \mbox{Tr}S_{w^k} (I - P_{w^k})\}. \eqno(19)$$
Taking the expectation and using (9), (12), (18), one obtains
$${\sf E}\min_{X}{\bar \lambda} (W, X)
\leq  4\epsilon + (M-1) \| {\bar S}_{\pi}^{\otimes n} P\|\mbox{Tr}\,
 {\sf E}P_{w},$$ for $n$ large enough, hence by (11), (17)
$$p(M, n)\leq  4\epsilon + (M-1)
2^{-n[\Delta H(\pi ) - 2\delta ]} $$
implying $C\geq {\bar C}$. Combined with the entropy bound, this gives

{\bf Theorem 2}. {\sl The capacity of the channel with $H(S_i)<\infty$ is given by}
 $$C = \max_{\pi} [H(\sum_{i\in
A}\pi_i S_i ) - \sum_{i\in A}\pi_i H(S_i )].\eqno(20)$$

For quasiclassical channel where the signal states are given by commuting density
operators $S_i$ one can use the classical bound
of Theorem 5.6.1 \cite{xgal} with
transition probabilities $S(\omega |i)$, where $S(\omega |i)$ are the
eigenvalues of $S_i$.  In terms of the
density operators it takes the form
$$
{\sf E}\min_{X}\bar{\lambda}(W, X) \leq \min_{0\leq s\leq 1}
 (M-1)^{s}\left(\mbox{Tr}\left[ \sum_{i\in A}
\pi_i S_i^{1 \over 1+s}\right] ^{1+s}\right) ^n.
\eqno(21)$$
The righthand side of (21) is meaningful for arbitrary
density operators, which gives a hope that this estimate could be
generalized to the noncommutative case (note that for pure states $S_i$
Theorem 1 gives twice the expression (21)).
This would not only give a
different proof of Theorem 2, but also a lower bound for the quantum
reliability function in the case of general signal states, eventually with
infinite entropy.
\vskip20pt
\centerline{\sc IV. Quantum channels with constrained inputs}
\vskip10pt
In classical information theory direct coding theorems for channels with
additive constraints are proved by using random coding with probability
distribution (8) modified with a factor concentrated on words, for which
the constraint holds close to the equality \cite{xgal}, Sec. 7.3.
The same tool can be applied to quantum channels \cite{xhol98}.
For definiteness
in this section we take for the input alphabet $A$ an arbitrary Borel subset
in a finite-dimensional Euclidean space $\cal E$.  We assume that the
channel is given by {\sl weakly continuous}
 mapping $x\rightarrow S_x$
from the input alphabet $A$ to the set of
density operators in $\cal H$. We assume that
a {\sl continuous} function $f$ on $\cal E$ is fixed and consider the
set ${\cal P}_1$ of probability measures $\pi$ on $A$ satisfying
$$
\int_{A} f(x) \pi (dx) \leq E .\eqno(22)$$

For arbitrary $\pi \in {\cal P}_1$ consider the quantity
$$
\Delta H (\pi ) = H ({\bar S}_{\pi}) - \int_A H (S_x ) \pi (dx) ,
\eqno(23)$$where
${\bar S}_{\pi} = \int_A S_x \pi (dx).$
Assuming the condition
$$\sup_{\pi\in{\cal P}_1} H( {\bar S}_{\pi}) < \infty ,\eqno(24)$$
we denote
$${\bar C} = \sup_{\pi \in {\cal P}_1} \Delta H (\pi ) .\eqno(25)$$

Let $p(M,n)$ denote the infimum of the average error probability over all
codes of size $M$ with words $w=(x_1,\ldots,x_n)$ satisfying
the additive constraint
$$f(x_1)+ \ldots + f(x_n)\leq nE .\eqno(26)$$

{\bf Theorem 3.} {\sl Under the condition (24) the capacity of the channel
with the input constraint (26) is given by (25), i. e.
$p(\mbox{e}^{n({\bar C}-\delta)}, n) \rightarrow 0$ ,
and
$ p( \mbox{e}^{n({\bar C}+\delta)}, n)\not\rightarrow 0$ for $\delta > 0$ as
$n\rightarrow\infty$.}

The proof uses the inequality (19) with the random coding modified as described
in \cite{xgal}, Sec. 7.3. The same method combined with the estimate (14) for
pure state channels gives lower bound  for the reliability function
modified with the factor const $\cdot$e$^{r[f(x) - E]}$, with $r\geq 0$.

Theorem 3, when applied to quantum memoryless 
Gaussian channels with the energy constraint
\cite{xhol98}, allows us to prove for the first time  their asymptotic
equivalence, in the sense of the information capacity, to the corresponding quasiclassical
``photon channels'', extensively studied from the origin of quantum
communications \cite{xgordon},
\cite{xle}, \cite{xcaves}. It is plausible that the 
equivalence extends also to waveform channels, in particular , that the 
infinite-band photon channel capacity \cite{xle}
$$C = \pi \sqrt{2 \over 3}\left[\sqrt{N+E \over \hbar} -
\sqrt{N \over \hbar}\right]$$
is equal to the properly defined
capacity of the quantum Gaussian channel
$$Y(t) = x (t) + Z(t) ; \qquad t\in [0,T],\quad T\rightarrow\infty ,$$ 
where $x (t)$ is the classical signal subject to the energy constraint
$\int_{0}^{T} x (t)^2 dt \leq E T ,$ and $Z(t)$ is the equilibrium quantum Gaussian noise
having the commutator
$$ [Z(t) , Z(s)] = i\hbar \int_0^{\infty}\omega\sin\omega (t-s) d\omega =
- i\hbar\pi\delta ' (t-s) ,$$ zero mean, and the correlation function
$ \langle Z(t) Z(s) \rangle = B_N (t-s) + K (t-s) ,$ with
$$B_N (t) = \hbar\int_0^{\infty}\omega {\cos\omega t \over \mbox{e}^{\beta\hbar\omega}
-1} d\omega = \hbar \, \mbox{Re}\sum_{k=1}^{\infty}{1 \over (k\beta\hbar + it)^2} , $$
$N$ and $\beta$ are related by $ N = B_N (0) = \hbar\pi^2 / 6\beta^2$,
and $$ K (t) = {\hbar \over 2}\int_0^{\infty}\omega\mbox{e}^{i\omega t}d\omega =
- {\hbar \over 2}[t^{-2} + i\pi\delta ' (t)]$$ is the zero temperature 
correlation $(B_0 (t) \equiv 0)$ .
However a complete proof is still lacking.
\newpage
{\sc V. Some further problems}
\vskip10pt
The present paper was entirely devoted to the ``classical-quantum'' channels,
in terminology of \cite{xhol77}),
and even in this case there are open problems, some of which were mentioned above.
Such channels can alternatively be described by (completely) positive maps
from noncommutative algebra of operators in $\cal H$ to commutative algebra
of functions on the input alphabet. More general ``quantum-quantum'' channels
are described by completely positive maps between noncommutative algebras.
The definition of capacity and the quantum entropy bound can be generalized
to this case \cite{xhol77}, \cite{xpetz}. However for such channels the new
difficult problem of optimization with respect to coding maps arises. In particular,
it is not yet known, whether the entropy bound optimized in this way is an
additive function on the product channel. An interesting preliminary
investigation of this situation is contained in the paper by Bennett, Fuchs and
Smolin \cite{xbennet}.

All these problems address transmission of classical information through
quantum channels. There is yet ``more quantum'' domain of problems
concerning reliable transmission of  entire quantum states under a given
fidelity criterion.  The very definition of the relevant ``quantum information''
is far from obvious. Important steps in this direction were made by Barnum,
Nielsen and Schumacher \cite{xbarnum} , who
in particular suggested a tentative converse of the relevant coding theorem. 
However the proof of the corresponding
direct theorem remains an open question.

\end{document}